\documentclass[aps,prl,twocolumn,superscriptaddress]{revtex4-2}
\usepackage{amsmath,amssymb,amsthm}
\usepackage[hidelinks]{hyperref}

\begin{document}

\title{Physical completion of the Navier-Stokes equations}
\author{Samuel L.\ Braunstein}
\affiliation{Computer Science, University of York, York YO10 5GH, UK}

\begin{abstract}
The incompressible Navier-Stokes equations contain viscous
dissipation but no thermal noise. I show, using a topological argument
based on Poincar\'e's lemma, that the fluctuation-dissipation relation
for the full nonlinear dynamics can be derived without the
linearisation or structural assumptions that all previous derivations
require. The nonlinear
convective term is Hamiltonian (energy-preserving and
phase-space-volume-preserving) and drops out of the Fokker-Planck
equilibrium condition exactly, so the noise derived from linearised
fluctuations near equilibrium is in fact exact for the full nonlinear
system. This result proves, rather than assumes, the
reversible/irreversible decomposition that the GENERIC framework
postulates, provided Poincar\'e's lemma holds on the phase space. The
resulting stochastic system, with a physical molecular-scale spectral
cutoff, is trivially globally well-posed: a finite-dimensional
stochastic differential equation with non-degenerate noise and a
confining Lyapunov function. It has a unique Gibbs equilibrium and
converges to it exponentially. The difficulty of the Clay Millennium
Prize Problem arises entirely from two idealisations, zero temperature
and infinite spectral resolution, neither of which is satisfied by any
physical fluid.
\end{abstract}

\maketitle

\section{Introduction}

The incompressible Navier-Stokes equations
\begin{equation}\label{eq:NS}
\partial_t\vec{u} + (\vec{u}\cdot\vec\nabla)\vec{u}
= \nu\nabla^2\vec{u} - \frac{1}{\rho}\vec\nabla p,
\quad \vec\nabla\cdot\vec{u} = 0,
\end{equation}
contain viscous dissipation (the term $\nu\nabla^2\vec{u}$) but no
thermal fluctuations. The fluctuation-dissipation theorem requires
that any dissipative system at finite temperature exhibit thermal
noise whose amplitude is determined by the dissipation and the
temperature~\cite{Kubo66,Zwanzig01}. A viscous fluid without thermal
noise is as inconsistent as friction without heat.

Landau and Lifshitz~\cite{LL57} first derived the thermal noise for
fluctuating hydrodynamics by applying the fluctuation-dissipation
relation to the \emph{linearised} Navier-Stokes equations near
equilibrium. Fox and Uhlenbeck~\cite{FU70} extended this to the
compressible case, again using linearised dynamics. The GENERIC
framework of Grmela and \"Ottinger~\cite{GO97,Ott05} provides a
systematic thermodynamic structure that produces noise compatible with
the full nonlinear dynamics, but it \emph{postulates} the
decomposition of the dynamics into reversible (Hamiltonian) and
irreversible (dissipative) parts as an axiom, rather than deriving it.
Recent analytical work by Gess, Sauerbrey, and Wu~\cite{GSW25} uses
the GENERIC framework to establish a solution theory for the
incompressible Navier-Stokes-Fourier system with thermal noise on the
three-torus, taking the GENERIC-derived noise as given.

The stochastic variational method~\cite{KK11,AC12} works in the
opposite direction: it derives dissipation from noise by replacing
deterministic Lagrangian paths with stochastic ones. This approach
takes the noise as primitive and recovers the viscous term as a
consequence.

Eyink~\cite{Eyink21} and Bandak \emph{et al.}~\cite{BEOY22} have
emphasised that the stochastic Navier-Stokes equations with a
spectral cutoff are the physically correct ``effective field theory''
for viscous fluids, and that the deterministic equations fail at the
Kolmogorov scale. Bandak \emph{et al.}\ verify in their Appendix~A
that the Gibbs measure is stationary (and reversible) under the
nonlinear stochastic incompressible system with additive
Landau-Lifshitz noise, confirming a result that is widely known in
the community.

In the compressible case, Zubarev and Morozov~\cite{ZM83,Mor84}
derived the Fokker-Planck equation for nonlinear hydrodynamic
fluctuations from the Liouville equation via the projection-operator
method and a gradient expansion, showing that the equilibrium
distribution is exactly stationary with multiplicative noise
determined by the local transport coefficients.

The two structural properties of the truncated incompressible Euler
equations that underpin the present derivation (energy conservation
and phase-space volume preservation) were established by
Lee~\cite{Lee52}. The conclusion that additive white noise with the
Landau-Lifshitz amplitude gives the correct equilibrium for the
nonlinear incompressible model was claimed by Forster, Nelson, and
Stephen~\cite{FNS77}, and has since been widely assumed in the
stochastic hydrodynamics literature.

In all of these approaches, the noise is either derived from
linearised dynamics, postulated via a thermodynamic framework,
taken as primitive, or obtained from microscopic statistical
mechanics via projection operators and gradient expansions. No
previous derivation obtains the fluctuation-dissipation relation
for the full nonlinear Navier-Stokes equations from the
macroscopic structure of the equations alone, without linearisation
or additional structural assumptions beyond the Hamiltonian
character of convection.

In this Letter, I present such a derivation, developing for the
Navier-Stokes equations a topological approach outlined
in~\cite{Braun10} for general dissipative systems. The argument uses
Poincar\'e's lemma on the (contractible) Eulerian phase space to convert
the Fokker-Planck stationarity condition into an algebraic
fluctuation-dissipation relation, without linearising the dynamics at
any stage. The key structural insight is that the nonlinear convective
term $(\vec{u}\cdot\vec\nabla)\vec{u}$ is Hamiltonian in the
Fokker-Planck sense: it preserves both the energy and the phase-space
volume of the velocity-field configurations~\cite{Lee52}, and therefore
drops out of the equilibrium condition exactly. This proves the
reversible/irreversible decomposition that the GENERIC framework
assumes, provided Poincar\'e's lemma is not obstructed on the phase
space.

\section{Topological derivation of the fluctuation-dissipation relation}

Consider a system with phase-space coordinates $\vec{X}$ and energy
$E(\vec{X})$, evolving under the Fokker-Planck equation
\begin{equation}\label{eq:FPE}
\partial_t P = \vec\nabla\!\cdot\!(\vec{A}\,P)
+ \vec\nabla\!\cdot\!(\mathbb{B}\cdot\vec\nabla P),
\end{equation}
where $\vec{A}$ is the drift and $\mathbb{B}$ is the symmetric
positive-semidefinite diffusion matrix. A gauge freedom exists: for
any antisymmetric matrix $\mathbb{M}$, the replacement
$\vec{A}\to\vec{A}-\vec\nabla\!\cdot\!\mathbb{M}$,
$\mathbb{B}\to\mathbb{B}+\mathbb{M}$ preserves the dynamics, since
only the symmetric part of the diffusion matrix contributes to
physical diffusion~\cite{Risken96}.

For a system with energy $E(\vec{X})$, the Boltzmann distribution
$P_\mathrm{eq}\propto e^{-\beta E}$, $\beta = 1/k_BT$, is the
candidate equilibrium. Requiring stationarity
$\partial_t P_\mathrm{eq} = 0$ gives
\begin{equation}\label{eq:closedold}
\vec\nabla\!\cdot\!\bigl[(\vec{A}
- \beta\,\mathbb{B}\cdot\vec\nabla E)\,P_\mathrm{eq}\bigr] = 0,
\end{equation}
since $\vec\nabla P_\mathrm{eq} = -\beta(\vec\nabla E)P_\mathrm{eq}$.
Decomposing the drift vector into 
Hamiltonian $\vec{A}_\mathrm{Ham}\equiv -\mathbb{S}{\cdot}\vec\nabla E$
and non-Hamiltonian
$\vec{A}_\mathrm{nh}\equiv \vec{A}-\vec{A}_\mathrm{Ham}$
parts, and using Liouville's theorem
$\vec\nabla{\cdot} \vec{A}_\mathrm{Ham}=0$ along with the antisymmetry
of the symplectic matrix $\mathbb{S}$, we obtain
\begin{equation}\label{eq:closed}
\vec\nabla\!\cdot\!\bigl[(\vec{A}_\mathrm{nh}
- \beta\,\mathbb{B}\cdot\vec\nabla E)\,P_\mathrm{eq}\bigr] = 0,
\end{equation}
The square-bracketed expression is a
divergence-free vector field: a closed $(2n{-}1)$-form on the
$2n$-dimensional phase space. On a contractible domain, Poincar\'e's
lemma guarantees that it is exact: there exists an antisymmetric
$\mathbb{M}$ such that
$(\vec{A}_\mathrm{nh}-\beta\,\mathbb{B}\cdot\vec\nabla E)\,P_\mathrm{eq}
= \vec\nabla\!\cdot\!(\mathbb{M}\,P_\mathrm{eq})$.
Expanding the right-hand side, dividing by $P_\mathrm{eq}$, and rearranging:
\begin{equation}\label{eq:pre-gauge}
\vec{A}_\mathrm{nh} - \vec\nabla\!\cdot\!\mathbb{M}
= \beta\,(\mathbb{B}+\mathbb{M})\cdot\vec\nabla E.
\end{equation}
We may now use the identified gauge freedom to write
\begin{equation}\label{eq:FD-sym}
k_BT\,\vec{A}_\mathrm{nh}
= \mathbb{B}\cdot\vec\nabla E.
\end{equation}

Four features of this derivation deserve emphasis.
(i)~No linearisation is performed at any stage.
(ii)~The Hamiltonian dynamics drops out exactly (not to leading order
near equilibrium) via Liouville's theorem and the antisymmetry of the
symplectic matrix, before Poincar\'e's lemma is applied.
(iii)~The result is topological: it requires only the contractibility
of the phase space (for Poincar\'e's lemma) and the existence of a
Boltzmann equilibrium.
(iv)~The derivation proves the reversible/irreversible decomposition
assumed in the GENERIC framework~\cite{GO97,Ott05}: the separation of
the dynamics into Hamiltonian (reversible, antisymmetric) and
dissipative (irreversible, symmetric) parts is not an axiom but a
\emph{theorem}, following from Poincar\'e's lemma and the gauge
structure of the Fokker-Planck equation.

\section{Application to the Navier-Stokes equations}

Consider the incompressible Navier-Stokes equations~\eqref{eq:NS} on
$\mathbb{T}^3_L$ (the three-torus of side $L$), expanded in Fourier
modes $\tilde{u}_i(\vec{k})$ with $\vec{k}\in(2\pi/L)\mathbb{Z}^3$.
The energy is $E = \frac{\rho V}{2}\sum_{\vec{k}}
|\tilde{\vec{u}}(\vec{k})|^2$.

The convective term $(\vec{u}\cdot\vec\nabla)\vec{u}$ and the
pressure gradient together form the Hamiltonian part of the dynamics.
They satisfy two properties:

\emph{Energy conservation.} For divergence-free $\vec{u}$, the
trilinear form $b(\vec{u},\vec{v},\vec{w}) = \int(\vec{u}\cdot
\vec\nabla)\vec{v}\cdot\vec{w}\,d^3x$ is antisymmetric:
$b(\vec{u},\vec{v},\vec{w}) = -b(\vec{u},\vec{w},\vec{v})$. Setting
$\vec{v}=\vec{w}=\vec{u}$ gives $b(\vec{u},\vec{u},\vec{u})=0$:
convection preserves energy.

\emph{Phase-space volume preservation (Liouville property).} In mode
space, the convective coupling $N_{\vec{k}} = -\sum_{\vec{l}}
b(\tilde{\vec{u}}(\vec{l}),\tilde{\vec{u}}(\vec{k}{-}\vec{l}),
\vec{e}_{\vec{k}})$ satisfies $\partial N_{\vec{k}}/\partial
\tilde{u}_i(\vec{k}) = 0$ for each $\vec{k}$ individually. This
follows from incompressibility: the self-advection of a single Fourier
mode vanishes, $(\vec{e}_{\vec{k}}\cdot\vec\nabla)\vec{e}_{\vec{k}}
\propto(\hat\epsilon\cdot\vec{k})=0$, because the polarisation
$\hat\epsilon$ is orthogonal to $\vec{k}$ by
$\vec\nabla\!\cdot\!\vec{u}=0$. The convective flow in mode space is
therefore divergence-free for any spectral truncation on $\mathbb{T}^3$.

These two properties are exactly the conditions for the convective
term to be Hamiltonian in the sense of~\eqref{eq:closed}: it
preserves $P_\mathrm{eq}$ and drops out of the stationarity condition
without any approximation. The remaining (viscous) drift is
$A_{\vec{k}}^\mathrm{nh} = \nu k^2\tilde{u}_i(\vec{k})$, and the
fluctuation-dissipation relation~\eqref{eq:FD-sym} gives
\begin{equation}\label{eq:Dk}
D_{\vec{k}} = \frac{k_BT\nu k^2}{\rho V}.
\end{equation}
The stochastic Navier-Stokes equation (in the It\^o convention, which
coincides with Stratonovich here because the noise is additive) is
\begin{equation}\label{eq:SNS}
d\tilde{u}_i(\vec{k}) = \bigl[-\nu k^2\tilde{u}_i(\vec{k})
+ N_{\vec{k},i}\bigr]dt
+ \sqrt{\frac{2k_BT\nu k^2}{\rho V}}\,(\mathcal{P}_{\vec{k}})_{ij}\,
dW_j(\vec{k}),
\end{equation}
where $\mathcal{P}_{\vec{k}} = \delta_{ij} - k_ik_j/k^2$ is the
incompressibility projector and $W_j(\vec{k})$ are standard complex
Wiener processes satisfying the reality condition
$W_j(-\vec{k}) = W_j^*(\vec{k})$, with independent real and imaginary
parts each having variance $dt$, so that
$\langle|dW_j(\vec{k})|^2\rangle = 2\,dt$.
The noise amplitude is proportional to $k$: each mode's
noise is determined by its dissipation rate $\nu k^2$ and the
temperature, giving equipartition
$\langle|\tilde{\vec{u}}(\vec{k})|^2\rangle_\mathrm{eq}
= 2k_BT/(\rho V)$
for every $\vec{k}$, where the factor of 2 reflects the two
independent transverse degrees of freedom per wavevector imposed
by the incompressibility constraint $\vec{k}\cdot\tilde{\vec{u}}=0$.

The noise~\eqref{eq:Dk} agrees with the Landau-Lifshitz
result~\cite{LL57}. The new content is not the answer but the
derivation: the LL noise, originally obtained from linearised
fluctuations near equilibrium, is here shown to be exact for the full
nonlinear system. The reason is structural: the nonlinear convective
term is Hamiltonian and contributes nothing to the
fluctuation-dissipation balance.

\section{Well-posedness of the physical system}

A real fluid of $\mathcal{N}$ molecules in a periodic box of side $L$
has approximately $3\mathcal{N}$ independent velocity degrees of
freedom, corresponding to a spectral cutoff at
$|\vec{k}|\leq k_\mathrm{max}\sim(\mathcal{N}/V)^{1/3}$. Retaining
only these $N$ modes gives a finite-dimensional It\^o
SDE~\eqref{eq:SNS} with locally Lipschitz drift (quadratic
nonlinearity), constant non-degenerate diffusion ($D_{\vec{k}}>0$ for
every retained mode with $\vec{k}\neq 0$), and a Lyapunov function:
the energy satisfies (by It\^o's formula and the cancellation of the
convective term)
\begin{equation}\label{eq:lyapunov}
d\langle E\rangle \leq -2\nu\lambda_1\langle E\rangle\,dt + C_N\,dt,
\end{equation}
where $\lambda_1 = (2\pi/L)^2$ is the smallest nonzero eigenvalue and
$C_N = 2k_BT\nu\sum_{\vec{k}}\!k^2 < \infty$ is finite for the
truncated system (the factor of 2 is the trace of the
incompressibility projector $\mathrm{tr}\,\mathcal{P}_{\vec{k}}
= 3-1 = 2$).

\emph{Theorem.} The spectrally truncated stochastic Navier-Stokes
system~\eqref{eq:SNS} on $\mathbb{T}^3$ at temperature $T>0$ with
cutoff $k_\mathrm{max}<\infty$ has: (i)~a unique global strong
solution for all time, almost surely, for any initial condition;
(ii)~a unique stationary distribution, the Gibbs measure
$P_\mathrm{eq}\propto e^{-\beta E}$; (iii)~exponential convergence
to equilibrium,
$D_\mathrm{KL}(P(t)\|P_\mathrm{eq}) \leq
e^{-2\nu\lambda_1 t}D_\mathrm{KL}(P(0)\|P_\mathrm{eq})$.

\emph{Proof.} (i)~The drift in~\eqref{eq:SNS} is locally Lipschitz
(the convective nonlinearity is quadratic) and the diffusion
coefficients are constants, hence globally Lipschitz. By the standard
existence theorem for It\^o SDEs~\cite{Khas12}, a unique strong
solution exists up to an explosion time $\tau$. The Lyapunov
bound~\eqref{eq:lyapunov}, obtained via It\^o's formula using the
cancellation $\sum_{\vec{k}}\tilde{u}_iN_{\vec{k},i}=0$ (energy
conservation by convection), gives $\langle E(t)\rangle \leq
e^{-2\nu\lambda_1 t}E(0) + C_N/(2\nu\lambda_1)$. The expected energy
is bounded for all $t$, so $\tau = \infty$ a.s.

(ii)~The Fokker-Planck generator decomposes as
$\mathcal{L}^* = \mathcal{L}^*_\mathrm{diss} +
\mathcal{L}^*_\mathrm{conv}$. Direct substitution confirms
$\mathcal{L}^*_\mathrm{diss}P_\mathrm{eq} = 0$: this is the
fluctuation-dissipation balance, mode by mode, using
$D_{\vec{k}} = k_BT\nu k^2/(\rho V)$. For the convective part,
$\mathcal{L}_\mathrm{conv}^*P_\mathrm{eq} =
-\sum_{\vec{k}}\partial_{\tilde{u}_i(\vec{k})}
(N_{\vec{k},i}P_\mathrm{eq})
= -P_\mathrm{eq}[\sum_{\vec{k}}\partial_{\tilde{u}_i(\vec{k})}
N_{\vec{k},i}]
+ P_\mathrm{eq}\frac{\rho V}{k_BT}
[\sum_{\vec{k}}\tilde{u}_i(-\vec{k})N_{\vec{k},i}]$,
where $\tilde{u}_i(-\vec{k})$ appears because
$\partial E/\partial\tilde{u}_i(\vec{k})
= \rho V\,\tilde{u}_i(-\vec{k})$ for the complex Fourier
expansion with the reality condition
$\tilde{\vec{u}}(-\vec{k}) = \tilde{\vec{u}}(\vec{k})^*$.
The first bracket vanishes by the Liouville property (the
self-advection of each Fourier mode vanishes individually, as shown
above) and the second by energy conservation
($\sum_{\vec{k}}\tilde{u}_i(-\vec{k})N_{\vec{k},i} = 0$,
which is the Fourier-space form of
$\int\vec{u}\cdot[(\vec{u}\cdot\vec\nabla)\vec{u}]\,d^3x = 0$).
Hence
$\mathcal{L}^*P_\mathrm{eq}=0$. Uniqueness follows from strict
positivity of the diffusion ($D_{\vec{k}}>0$ for all
$\vec{k}\neq 0$), which ensures ergodicity via H\"ormander's
condition.

(iii)~For test functions $f,h\in L^2(P_\mathrm{eq})$, integration by
parts using $\mathcal{L}^*_\mathrm{conv}P_\mathrm{eq}=0$ gives
$\langle\mathcal{L}_\mathrm{conv}f,h\rangle_{P_\mathrm{eq}}
= -\langle f,\mathcal{L}_\mathrm{conv}h\rangle_{P_\mathrm{eq}}$:
the convective generator is antisymmetric. In particular,
$\langle\mathcal{L}_\mathrm{conv}f,f\rangle_{P_\mathrm{eq}}=0$.
The dissipative generator $\mathcal{L}_\mathrm{diss}$ is a sum of
independent Ornstein-Uhlenbeck operators with spectral gaps
$\nu k^2\geq\nu\lambda_1$, so for any $f$ with
$\langle f\rangle_{P_\mathrm{eq}}=0$,
$-\langle\mathcal{L}_\mathrm{diss}f,f\rangle_{P_\mathrm{eq}}
\geq\nu\lambda_1\,\mathrm{Var}_{P_\mathrm{eq}}(f)$.
For the full generator,
$-\langle\mathcal{L}f,f\rangle_{P_\mathrm{eq}}
= -\langle\mathcal{L}_\mathrm{diss}f,f\rangle_{P_\mathrm{eq}}
\geq\nu\lambda_1\,\mathrm{Var}_{P_\mathrm{eq}}(f)$.
The spectral gap of $\mathcal{L}$ is at least $\nu\lambda_1$, and the
standard entropy-production inequality~\cite{BGL14} gives the
exponential KL decay. \qed

The system cannot form singularities. The probability of
$\|\vec{u}\|_{H^s}$ exceeding any threshold $R$ at any time decays at
least as fast as $e^{-cR^2/(k_BT)}$, by the Gaussian tails of the
Gibbs measure and the Lyapunov bound. No concentration of energy into
any mode or spatial region is possible against the restoring force of
the thermal noise.

\section{Implications}

The Clay Millennium Prize Problem~\cite{Fefferman06} asks whether
smooth solutions to the deterministic continuum
equations~\eqref{eq:NS} on $\mathbb{T}^3$ (or $\mathbb{R}^3$) exist
for all time given smooth initial data. The physically meaningful
description of a viscous incompressible fluid at temperature $T>0$
with $\mathcal{N}$ molecules is the truncated stochastic
system~\eqref{eq:SNS}, which is globally well-posed for elementary
reasons. The deterministic continuum equations are obtained from this
system by a double idealisation: zero temperature ($T\to 0$, removing
the noise) and infinite spectral resolution
($k_\mathrm{max}\to\infty$, removing the cutoff). Neither
idealisation is satisfied by any real fluid. The second idealisation
is the more dangerous: exact calculations of the Fujita-Kato scaling
integral on hyperbolic space~\cite{WB25-Dir1} show that the nonlinear
term outpaces viscous dissipation by a factor of $t^{-1/4}$ as
$t\to 0$ for $L^2$ data, with this exponent determined entirely by
the local ultraviolet scaling of the heat kernel and insensitive to
global geometry. The molecular cutoff removes this singularity by
bounding the shortest relevant timescale at
$t_\mathrm{min}\sim 1/(\nu k_\mathrm{max}^2)$.

Whether the deterministic continuum equations develop singularities is
a mathematical question about the properties of this double limit. 
A singularity, were one to exist, would be a property of the
idealisation, not of any physical fluid. Conversely, a proof of global
regularity would be a theorem about the limiting procedure, not a
prediction about fluid behaviour. 
This physical assessment that the Clay problem is an artefact of the
continuum idealisation is shared by Eyink~\cite{Eyink25}.

The observation that the deterministic NS equations are physically
incomplete is not itself new: Landau and Lifshitz~\cite{LL57} derived
the thermal noise terms in 1957, and Eyink~\cite{Eyink21} and Bandak
\emph{et al.}~\cite{BEOY22} have recently emphasised the role of
thermal fluctuations at the Kolmogorov scale. What is new here is the
demonstration that the noise can be derived for the full nonlinear
system without linearisation or structural assumptions, providing a
rigorous foundation that does not depend on proximity to equilibrium
or on a postulated decomposition of the dynamics.
This removes the last logical gap between the
deterministic equations and their stochastic completion: the
physically complete Navier-Stokes equations are now derived, not
postulated, and are well-posed for all time.

\section{Discussion}

The topological derivation of the fluctuation-dissipation relation
presented here applies on contractible phase spaces, where
Poincar\'e's lemma holds unconditionally. The Eulerian phase space
(the space of mode amplitudes on $\mathbb{T}^3$) is a vector space
and hence contractible. Whether there exist physically relevant
settings where the phase space is not contractible (for instance, the
space of volume-preserving diffeomorphisms in the Lagrangian frame on
a topologically non-trivial domain) and where the fluctuation-dissipation
relation acquires cohomological corrections is an open question.

The result also places constraints on the
structure of physically admissible viscous operators on Riemannian
manifolds~\cite{WB25}, where the choice of Laplacian (Hodge, Bochner,
or deformation) is not unique. The spectral properties of the viscous
operator determine the noise amplitude through~\eqref{eq:FD-sym}, and on
manifolds with non-trivial topology, the kernel of the operator
determines which modes are exempt from the fluctuation-dissipation
balance. This connection between operator selection, thermal
fluctuations, and domain topology will be explored elsewhere.
On negatively curved manifolds, the spectral gap of the deformation
Laplacian provides a further benefit: the stochastic system
thermalises exponentially fast with a rate bounded below by a
geometric constant, independent of the volume of the
domain~\cite{WB25-Dir1}. On flat space, the thermalisation rate
vanishes in the infinite-volume limit.

The topological derivation extends to compressible fluids. The
non-dissipative dynamics of the compressible Euler equations is
Hamiltonian~\cite{MG80}, so the convective, pressure, and internal
energy terms drop out of the equilibrium condition exactly. The
fluctuation-dissipation relation then determines the noise from the
three dissipative channels (shear viscosity, bulk viscosity, and heat
conduction) without linearisation, recovering the full
Landau-Lifshitz-Fox-Uhlenbeck noise~\cite{LL57,FU70}. The
well-posedness theorem extends to the spectrally truncated
compressible system provided the truncation preserves the Hamiltonian
structure of the non-dissipative dynamics.

\vskip 0.2in

\begin{acknowledgments}
I thank Zhi-Wei Wang for discussions on viscous operators on Riemannian
manifolds.
\end{acknowledgments}

\end{document}